\documentclass[epj,final]{svjour}[20-10-2011]
\usepackage{amssymb}
\usepackage{amsmath}
\usepackage{graphicx}
\usepackage{hyperref}

\begin{document}

\title{Layers of Cold Dipolar Molecules in the Harmonic Approximation}

\author{J.~R. Armstrong, N.~T. Zinner, D.~V. Fedorov and A.~S. Jensen}
  \institute{Department of Physics and Astronomy - Aarhus University, Ny Munkegade, bygn. 1520, DK-8000 \AA rhus C, Denmark}
\date{\today}

\abstract{
We consider the $N$-body problem in a layered geometry containing 
cold polar molecules with dipole moments that are polarized perpendicular
to the layers. A harmonic approximation is used to simplify the 
Hamiltonian and bound state properties of the two-body 
inter-layer dipolar potential are used to adjust this effective interaction.
To model the intra-layer repulsion of the polar molecules, we introduce 
a repulsive inter-molecule harmonic potential and discuss how its strength can 
be related to the real dipolar potential. However, to explore different structures 
with more than one molecule in each layer, we treat the 
repulsive harmonic strength as an independent variable in the problem.
Single chains containing one molecule in each layer, as well as 
multi-chain structures in many layers are discussed and their
energies and radii determined. We extract the normal modes of 
the various systems as measures of their volatility and eventually
of instability, and compare our findings to the excitations in 
crystals. We find modes that can
be classified as either chains vibrating in phase or as layers
vibrating against each other. The former correspond to acoustic 
and the latter to optical phonons. For the acoustic modes, our
model predicts a smaller sound speed than one would naively
get from expansion of the dipolar potential to second order
around the origin.
Instabilities can occur for 
large intra-layer repulsion and produce diverging amplitudes of
molecules in the outer layers, and our model predicts how the 
breakup takes places.
Lastly, we consider experimentally
relevant regimes to observe the structures. The harmonic 
model considerd here predicts that for the multi-layer systems under current 
study chains with one molecule in
each layer are always bound whereas two chains comprised of two 
molecules in each layer will not be bound. However, since realistic 
systems have external confinement prevention the molecules
from escaping to infinity, we still expect the unstable 
modes to show up as resonances in the dynamics.
}
\authorrunning{J.~R. Armstrong {\it et al.}}
\titlerunning{Layers of Cold Polar Molecules in the Harmonic Approximation}
\maketitle

\section{Introduction}
The experimental study of cold dipolar molecules is a rapidly
accelerating field \cite{dipoleexp1,dipoleexp2,dipoleexp3,dipoleexp4,dipoleexp5,dipoleexp6,ni2010,ospelkaus2010,carr2009,miranda2011}.
The long-range anisotropic forces of polar molecules can be controlled by external alignment and 
tuned to be both repulsive and attractive depending on the geometric setup, which 
has lead to a significant amount of interesting theoretical proposals \cite{tr1,tr2}.
In particular,
stacks of thin layers containing polarized
molecules interacting via dipole-dipole potentials have been suggested \cite{2dt1,2dt2,2dt3,2dt4,wang2006} 
as a way to control
the losses that can severely influence three-dimensional experiments with strong dipolar
forces \cite{ni2010,ospelkaus2010,lushnikov2002,micheli2010}. The layered structure and 
the long-range nature of the interactions holds promise for the realization of 
interesting few- \cite{fewbody1,fewbody2,fewbody3,fewbody4,fewbody5,cremon2010,armstrong2010,artem2011-1,artem2011-2,zinner2011a} and many-body states
\cite{mb1,mb2,mb3,mb4,mb5,mb6,mb7,mb8,mb9,mb10,mb11,mb12}.
Recently, a stack of layers have been realized with fermionic polar $^{40}$K$^{87}$Rb molecules \cite{miranda2011}
and further cooling should produce some of the novel phases that have been proposed.

An interesting question to pose for layered systems of dipolar molecules is 
related to their tendency for crystallization when the dipole moment becomes
large. This phenomena is similar to the famous Wigner crystal phase of 
the electron gas \cite{wigner1934-1,wigner1934-2}. However, whereas the 
electron gas crystallizes at low density where the Coulomb 
interaction dominates, a system of dipoles will crystallize at high density
where the kinetic energy is negligible compared to the dipole force. In a
single two-dimensional (2D) layer with dipoles 
oriented perpendicular to the layer this has been studied in classical 
simulations \cite{kalia1981-1,kalia1981-2}, and more recently in quantum 
Monte Carlo simulations \cite{dc1,dc2,dc3}, and for 
large dipole moments evidence for a triangular crystal structure is
found. Studies in one-dimensional (1D) dipolar systems also find
interesting crystal phases \cite{1dc1,1dc2,1dc3,1dc4,1dc5,1dc6,1dc7,1dc8} 
which are similar to those seen in
ion Coulomb crystals \cite{fishman2008-1,fishman2008-2}. Crystal phases
have also been found in 2D with an in-plane optical lattice for arbitrary
polarization \cite{quin2009-1,quin2009-2}.

Here we are interested in the case of a multi-layer system with 
dipoles perpendicular to the planes as in recent experiments \cite{miranda2011}.
The possibility of having bound states consisting of chains with one 
molecule in each layer has been discussed for both 
bosonic \cite{wang2006,wang2008a} and fermionic molecules \cite{santos2010}.
In the limit of large dipole moments, a study of the two-layer (bilayer) 
case using classical dynamics showed that the triangular crystals appear
spatially correlated, i.e. the molecules in the crystal lattice sit 
on top of each other \cite{lu2008}. This is also the expected behaviour for 
more than two layers and we note that a recent quantum Monte Carlo study
finds evidence of a transition to rough chains for bosonic molecules at 
a critical dipole strength \cite{barbara2011}.

In the present work we consider strongly interacting
distinguishable dipolar molecules before they enter the crystal phase. Rather
than fixing the position of the molecules to a particular
lattice and calculate fluctuations about this state, we consider an
$N$-body system with a fixed number of molecules in each layer
and study the quantum spectrum to obtain information about the 
modes of the system before it crystallizes.
Most of the studies on multi-layer systems mentioned above are 
based in one way or another on a harmonic approximation to the 
interaction potential which renders the otherwise intractable $N$-body
problem solvable in certain limits, and provides ground state and 
excitation modes of the system. In this paper we also make use of the 
harmonic approximation to derive a solvable $N$-body problem. 
However, we go beyond previous studies in two important aspects; 
i) we use the two-body binding energy to fix the interlayer attractive
interactions for the effective harmonic hamiltonian and ii) 
we include parametrically the repulsive interaction between 
molecules in the same layer which has been neglected thus far 
except in the classical calculations of Ref. \cite{lu2008}.

To arrive at our effective harmonic Hamiltonian, we use the 
method recently formulated in Refs. \cite{armstrong2010,armstrong2011}.  
We proceed in two steps, that is first the
one- and two-body interactions are approximated by quadratic forms,
and second the resulting $N$-body Schr\"{o}dinger equation is solved
exactly.  We replace the true interactions with quadratic forms in the
molecule coordinates, either by direct fits of the potentials or by
adjusting parameters to reproduce crucial properties.  This
approximation allows analytical investigations of the $N$-body system
with the properties expressed in terms of the two-body
characteristics.  

The harmonic model studied here predicts that chains with a single molecule
are stable for any interaction strength. A complex with two molecules in 
each layer, i.e. two chains in close proximity, will most likely not be 
bound as we estimate the critical strength for breakup to be below the 
strength in current experimental setups. However, experiments are always
performed in the presence of an external confinement and we expect the 
unstable modes of complexes with multiple chains to appear as resonances.
We therefore study the modes as function of the intralayer
repulsion to determine how the system breaks into single chains and 
elucidate its structure.

The paper is organized as follows. In Section \ref{methodsect} we describe the 
harmonic approximation method used and Section \ref{oneptclsect}
discusses in detail the chain structure with one molecule in each
layer. This involves energies, radii, and normal mode excitations.
Section \ref{multsect} deals with multiple chains and layers with many molecules.
The intralayer repulsion is taken as a parameter in the harmonic approximation.  
The normal modes reveal the most likely decay mechanism, and extract configurations of the
most unstable degrees of freedom. In particular, we calculate the critical stability properties
of the system and discuss how the system breaks up into smaller structures.
In Section \ref{denssect} we discuss how the
intralayer repulsion influences the system size and densities, and we discuss relative
energies of the many-body systems. Finally, Section \ref{conclusionsect} contains summary and conclusions.

\begin{figure}\centering
\includegraphics[width=0.5\textwidth]{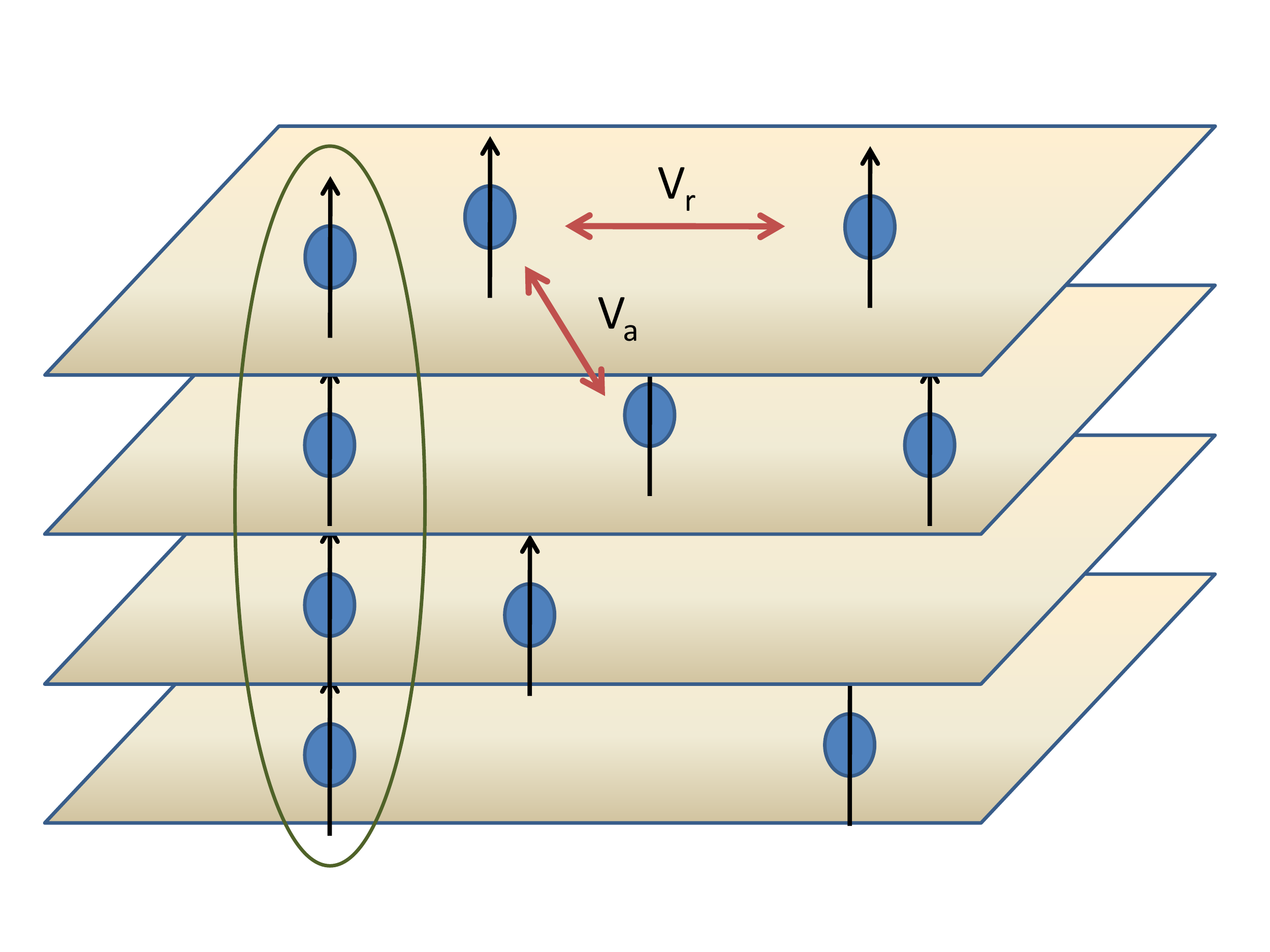}
\caption{Pictorial view of the multi-layered system containing dipolar molecules. The attractive interlayer ($V_a$) and 
repulsive intralayer ($V_r$) are illustrated, along with a potential bound chain state containing four molecules.}
\label{pictorial}
\end{figure}

\section{Method}\label{methodsect}
We consider a system of $N$ dipolar molecules of mass, $m$, and dipole
moment, $D$, distributed in a series of parallel two-dimensional
layers separated by a distance $d$ as illustrated in figure \ref{pictorial}.  
The planar coordinates of the
k'th molecule are $(x_{k},y_{k})$.  The dipoles are oriented
perpendicular to the layers and the molecules interact pairwise
through the dipole-dipole potential, $V$, which in different planes
is given as
\begin{eqnarray}
V(x,y,n) = D^2 \frac{x^2+y^2 - 2(nd)^2}{(x^2+y^2+(nd)^2)^{5/2}},
\label{e30}
\end{eqnarray}
where $n$ is the number of layers separating the two molecules, and
$(x,y)$ are the relative coordinates of the two molecules.  
We define the dimensionless strength, $U=mD^2/(\hbar^2d)$. For $n=1$
we have a nearest neighbour interaction, for $n=2$ a next-nearest 
neighbour interaction and so forth. Below 
we will also use the standard notation for the radius $r=\sqrt{x^2+y^2}$. 
Molecules
in the same layer repel each other and if identical are also subject to
(anti)symmetrization conditions for fermions and bosons. Here we will 
ignore symmetry considerations and consider our molecules distinguishable.
Since we work in the strongly-coupled regime and we do not expect the chains
to have large overlap this is a reasonable assumption.
The Hamiltonian for the $N$-body system is then
\begin{equation}
H=-\frac{\hbar^2}{2m} \sum_{k=1}^N
\left(\frac{\partial^2}{\partial x_k^2}+\frac{\partial^2}{\partial y_k^2}\right)
 +\frac{1}{2}\sum_{i\neq k}V(x_{ik},y_{ik},n_{ik}) ,
\label{hamil}
\end{equation}
where $(x_{ik},y_{ik})=(x_{i}-x_{k},y_{i}-y_{k})$, and $V$ for
different layers is from equation (\ref{e30}) or the corresponding repulsion
for molecules in the same layer. This $N$-body Hamiltonian is solved using the 
method described in Ref.~\cite{armstrong2011}. Note that there is no external 
trapping potential in any of the planes. The $N$-body structures we consider
are self-bound, i.e. they are held together by the attractive interactions
between the layers. For configurations that have more than one molecules
in a single layer, the attractive forces will have to overcome the 
repulsive intra-layer interactions. This will be disucssed in great detail later
and will give raise to critical values of this intra-layer repulsion above 
which there are no bound structures.

\begin{figure}\centering
\includegraphics[width=0.5\textwidth]{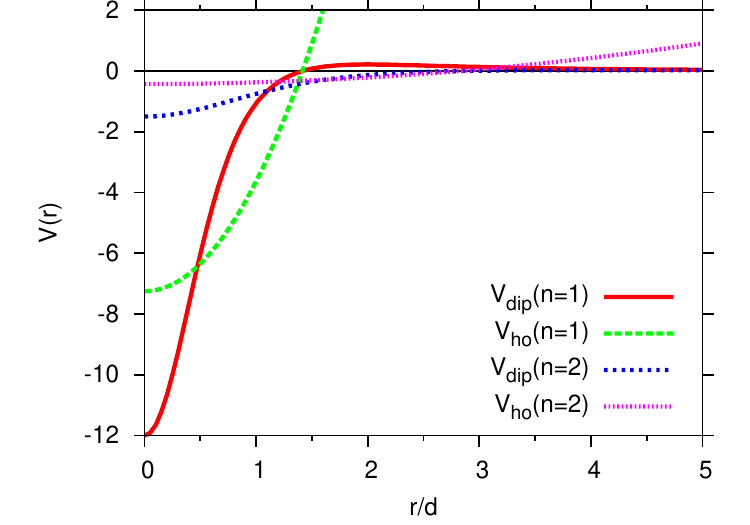}
\caption{Potential energy curves for the nearest neighbour 
  dipole-dipole interaction ($V_{dip}(n=1)$) and
  the harmonic oscillator potential ($V_{ho}(n=1)$) with the same binding energy.  
  The next-nearest neighbour dipole-dipole interaction is also shown ($V_{dip}(n=2)$) as 
  well as the oscillator potential with the matching binding energy ($V_{ho}(n=2)$).  
  The dipole strength is here chosen as $U=6$.
  The resulting binding energies are $B_2 = 3.44$ and $0.20$
  in units of $\hbar^2/md^2$.  }
\label{potplot}
\end{figure}   

To render properties of systems with $N$ molecules analytically 
tractable, we simulate the effects of the
dipole-dipole interaction by harmonic oscillator potentials as
described in \cite{armstrong2011}.  We replace the actual interaction in
equation (\ref{e30}) with a shifted harmonic oscillator, $V_{ho}$, which has
its node in the same place as the dipole-dipole potential, i.e.,
\begin{eqnarray}
V_{ho}(x,y,n) = V_0\left(\frac{x^2+y^2}{2(nd)^2}-1\right).
\label{HOpot}
\end{eqnarray}               
The potentials in equations (\ref{e30}) and (\ref{HOpot}) are negative (and
positive) in the same regions, and have nodes in the same places, see
figure \ref{potplot}.  The strength, $V_0$, is finally adjusted to
reproduce the two-body dipole-dipole binding energy.  This means that
$V_0$ is a specific function of the dipole moment, the molecule mass, and
distance between the layers, $V_{0}(D^2,m,d)$  The corresponding oscillator frequency,
$\omega_a$, is then given by
\begin{equation} \label{e40}
\omega_{a}^2=\frac{2V_{0}(D^2,m,d)}{m(nd)^2}\;,
\end{equation}
where the reduced mass of the two-body system is $m/2$.

\begin{figure}\centering
\includegraphics[width=0.5\textwidth]{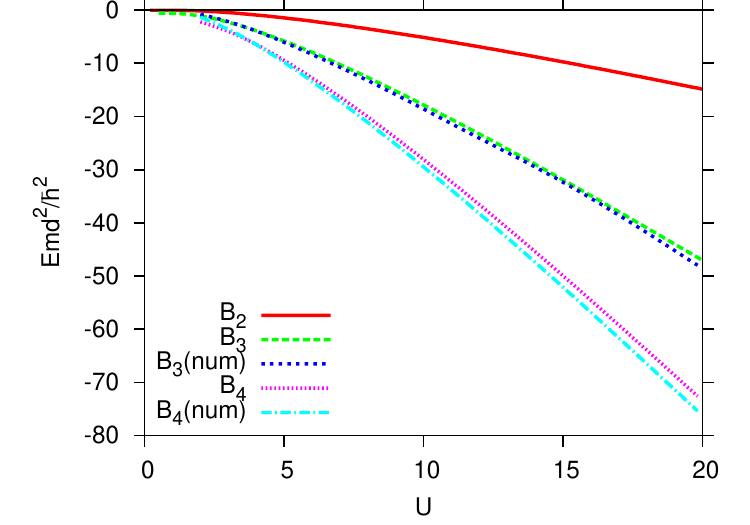}
\caption{Binding energies for two, $B_2$, three $B_3$, and four $B_4$ molecules as
  functions of strength, $U=mD^2/\hbar^2 d$, of the dipole potential.  The
  oscillator approximation (solid, red) and the accurate numerical
  calculation (dashed, blue) are compared for three ($B_3(\text{num})$) and four molecules
  ($B_4(\text{num})$). The numerical calculations are courtesy of A.~G. Volosniev using the 
  method described in \cite{artem2011-1,artem2011-2}.}
\label{endip}
\end{figure}   

The key ingredient is here the two-body binding energy, $B_2$, which
from equation (\ref{hamil}) for two molecules is seen to scale with layer
distance, $nd$, as
\begin{equation} \label{e50}
B_2(n) =  \frac{\hbar^2}{m (nd)^2} f\bigg( \frac{m D^2}{nd \hbar^2} \bigg)  \;,
\end{equation}
where $f$ is a universal function.  For $n=1$, that is, for neighbouring
layers, we show $B_2$ in figure \ref{endip} as function of potential
strength, $D^2$. The two-body interaction for pairs of molecules in
layers separated by the distance, $nd$, is then obtained through the
scaling in equation (\ref{e50}). Upon comparing the root-mean-square
radius of the exact solution to the oscillator approximation one finds 
very good agreement for $U\gtrsim 2$ which is also where
the binding energy of the exact and harmonic approximation for 
two molecules are very close \cite{armstrong2011,zinner2011a}.

These energies are then the crucial input in the $N$-body
calculations. At small strength the approach to zero is extremely
strong \cite{artem2011-1,artem2011-2} whereas at larger strengths the behaviour is smooth
and in a region almost linear in $U$.  We need two-body interactions
between each pair, the most important being between adjacent layers at
a distance $d$.  For larger than one-layer separations the potential
becomes increasingly more shallow, flat and spatially extended, see
figure \ref{potplot}.  The scaling of $B_2$ with layer number is
$n^{-2}$ computed for a strength $U/n$ (see equation (\ref{e50})).

This procedure of adjusting strengths for different shapes of 
potentials to get the same binding energy has proved efficient in
other contexts where effective interactions are used.  The method is
frequently applied in dripline nuclear physics, where knowledge about the
systems can be limited to an energy of one bound state, or resonance
\cite{jen04}.  It is also the philosophy employed for cold atoms where
the scattering length is used as the parameter characterizing a short-range
two-body interaction, and subsequently used in $N$-body calculations.
However, the dipolar interaction is long-ranged. In fact, for 
typical experimental densities, the range of the dipole-dipole forces
is comparable to the inter-particle distance. This implies that one
should take more than just the zero-energy properties into account. 

The accuracy of the procedure is tested by comparing the oscillator
approximation with numerical calculations of the energies of three
and four molecules with one in each layer using a method similar to the 
one described in \cite{artem2011-1,artem2011-2}, see
figure \ref{endip}.  The remarkable precision of the oscillator results
for three molecules is reassuring in the applications on $N$ molecules.

We note that the procedure used here to fix the frequency of the 
attractive interlayer interaction is naturally very different from 
expanding the potential in equation (\ref{e30}) for $r/d\ll 1$ to 
get the harmonic term \cite{zinner2011a}. We include more accurately the two-body 
properties into the $N$-body system and expect therefore to provide
a better approximation for the dynamics of a single chain and also 
for the interaction of several chains as we discuss below.

\subsection{Repulsive In-Plane Interactions}
Two polar molecules in the same layer repel 
each other when both dipoles are
perpendicular to the plane. This repulsive intralayer interaction 
scales with the dipole strength similarly to the 
attractive interlayer dipolar interaction. Within the framework 
of effective harmonic hamiltonian we would like to replace the
real intralayer potential by something that can model a repulsion, 
which can be achieved by using an inverted oscillator.
The repulsive interaction is therefore
replaced by a term of the form
\begin{equation} \label{e60}
V_{rep}(x,y)=-\frac{1}{4} m \omega_r^2\left(x^2+y^2\right)+V_{r0},
\end{equation}
where $1/4$ arise from use of the reduced mass, $\omega_r$ is the
corresponding frequency, and $V_{r0}$ is a constant shift.  This
inverted oscillator favors a large (infinite) distance between the
molecules, and acts consequently as a repulsion. 
The shift is parametrized as
\begin{equation}
V_{r0}= \frac{1}{4} m \alpha^2 \omega_r^2 d^2,
\label{repshift}
\end{equation}   
where we use $d$ as the natural length, and leave the dimensionless
scale factor, $\alpha$, for later adjustment.  The distance where the
potential changes sign is then $r=\alpha d$.  

We will demonstrate below that the inverted oscillator
provides us with a parametrically sound way of including repulsive
forces in our model.
That is, it provides a handle on the effect 
of repulsion in the detailed balance between repulsively and 
attractively interacting pairs of molecules. Therefore we are 
able to consider various properties of several chains in multiple
layers which is one of the goals of this work.

At this point we must caution that one might naively try to 
fit the frequency of the repulsive term above to the true 
intralayer dipole interaction which has the form $V(r)=\lambda/r^3$.
Here we introduce the dipolar strength $\lambda$ since we want to 
separate it from the strength of the attractive interlayer potential
(\ref{e30}) for later discussion. 
If we have a system at a certain density, $n$, then the characteristic
energy of the true repulsion can be estimated as $\lambda n^{3/2}$. 
The energy scale of an oscillator is usually $\hbar\omega_r$ and 
we then arrive at 
\begin{align}\label{ocrit}
\frac{md^2\omega_r}{\hbar}=\frac{m\lambda}{\hbar^2 d} (nd^2)^{3/2}.
\end{align}
For current experiments \cite{miranda2011} with $m\lambda/\hbar^2 d\sim 0.1$ and 
$nd^2\sim 1$, this gives a very small contribution from the repulsive 
term. However, the procedure of harmonic approximation breaks down 
for small $U$ and/or $\lambda$, and both the attractive and repulsive parts of the 
potential cannot be modelled by an oscillator in this limit. Only in the 
limit of large $U$ does the approximation become valid for the 
single-chain structures as shown above. Quantitatively, we expect that 
harmonic oscillator wave functions are reliable for 
$U\gtrsim 2$ \cite{armstrong2010,zinner2011a}.
Below this value, the binding energy decreases rapidly and the state becomes
extended \cite{artem2011-1,artem2011-2}.

As will be shown below, there
are critical values of $\omega_r(U)$ above which the system becomes unstable
as indicated by the normal mode frequencies that become imaginary.
If we take densities $nd^2\sim 1$, the $\omega_r$ given by equation (\ref{ocrit})
is always larger than the critical frequency 
and the system is unstable. Furthermore, the densities
we calculate below for the system are much larger than $nd^2=1$, so this 
gives an apparent inconsistency. 

There are, however, two major points that 
invalidate the identification in equation (\ref{ocrit}). First,
there is no zero-point energy scale for an inverted oscillator, and 
using $\hbar\omega_r$ as such is meaningless. One might attempt to 
make sure that the force (gradient of the potential) of the true
potential and the inverted oscillator match in a region around zero, but
this suffers from the same inconsistencies. The second, and more 
important objection, is that we must choose the repulsion in a 
manner that is consistent with the philosophy of using effective 
harmonic approximations already employed for the attractive 
interlayer interaction discussed above. There we really use the 
knowledge of two-body states in the system to fix a potential that 
reproduces features such as energy and potential shape.

In order to apply our philosophy, we can
compare to small systems that have the effect of the repulsion 
included, but which can be solved by other means. The obvious 
choice is the four-body state with two molecules in each of two layers 
or two molecules in one layer and one molecule in each of the two
adjacent layers in a setup with three layers. 
We could then compare the energetics and fit the strength of the 
repulsive term. Exact results on these configurations have been
reported very recently in Ref.~\cite{artem2011c}. These results
indicate that such systems are actually unstable for a 
critical intralayer repulsion that is much smaller than the
situation where attraction and repulsion have the same value 
($U=\lambda$ below). Ref.~\cite{artem2011c} also 
find that longer chains (five layers or more) 
could bind extra molecules as long as the additional ones
are not in adjacent layers. This hints at a competition 
between attractive and repulsive interaction terms that is conveniently
described within the harmonic oscillator approach as we 
demonstrate below. The results of Ref.~\cite{artem2011c}
are, however, difficult to use as a fit for the repulsive
frequency in the harmonic model due to the instability of
the states. We therefore follow a different line
of argument which relates the number of bound states in 
the (inverted) repulsive potentials as we now explain. 

In a single plane, the pure inverse cubic dipolar repulsion 
obviously has no two-body bound states. However, if we introduce an external trapping
potential, i.e. an external oscillator to keep the molecules confined, 
then the two-body spectrum would be shifted upward due to the 
repulsion. One could then imagine adding a repulsive oscillator term
to fix the two parameters of the potential in (\ref{e60}) so 
that, say, the two lowest bound state energies match (or the 
ground state and the root-mean-square radius). Unfortunately,
the results of such a procedure depend strongly on the 
choice of external oscillator potential. If one relates 
this external confinement to the density of particles in the
layer (in a manner similar to (\ref{ocrit})), then we are left
with a two-body interaction that depends on the properties
of the entire system. We consider this situation extremely
inconvenient. This dependence of two-body physics on the 
many-body problem is also inherent in the naive estimate 
in (\ref{ocrit}).

What can be done instead, is to consider the number of bound 
states that the {\it inverse} dipolar potential, i.e. an 
{\it attractive} potential, allows and compare this to $-V_{rep}$
(\ref{e60}) (a normal oscillator that is shifted below zero at 
the origin). We consider only those bound states that have 
negative energy even though the shifted normal oscillator will
of course have the usual ladder spectrum of states. This 
explains the upper limit in (\ref{intrep}) below.
We are interested
in the stability of structures with several molecules in 
each layer, and therefore we need only require that the
repulsion be reproduced within the range where there 
is an attraction to particles in other layers. This means that 
the shift term in (\ref{repshift}) should ensure that the 
inverted oscillator crosses zero along the outer repulsive 
barrier of (\ref{e30}). This is roughly in the region
$\sqrt{2}\leq \alpha \leq 2\sqrt{2}$ (the upper bound 
being defined as the point at which the repulsion 
of (\ref{e30}) is only 1\% of its attractive value at the 
origin).
The parameter $\alpha$ is therefore still somewhat arbitrary 
although we will compute a suitable value based on the 
energetics of barely stable configurations of molecules in
section \ref{denssect}.

In a realistic setup, the layers have finite width, $w$, and therefore
the $1/r^3$ repulsion is modified at small distances \cite{cremon2010}
and attains a logarithmic dependence on relative distance of two
molecules. The crossover happens for distances of order the layer
width (typically $w\sim 0.1d-0.2d$ in experiments). Matching at $w$, 
we have
\begin{equation}\label{realdip}
V_{dip}(r)=\left\{
\begin{matrix}\frac{\lambda}{r^3} &, & r>w \\
-3\frac{\lambda}{w^3}\ln\left(\frac{r}{we^{1/3}}\right)&,& r<w \end{matrix}
\right. .
\end{equation}

An estimate for the number of bound states in a two-dimensional potential can 
be found by integrating the (absolute value of the) negative part of the potential 
over space \cite{khuri2002}. For the dipolar potential we find
\begin{equation}
\int_{0}^{\infty}V_{dip}(r)r \,dr=\frac{9}{4}\frac{\lambda}{w},
\end{equation}
while for the oscillator of (\ref{e60}) we find
\begin{equation}\label{intrep}
\int_{0}^{\alpha d}V_{rep}(r)r \,dr=\frac{1}{16}m\omega_{r}^{2}d^4\alpha^4.
\end{equation}
If we equate these expressions we obtain a relation between the 
repulsive frequency, $\omega_r$, and the parameters $\lambda$, 
$\alpha$, and $w$, which reads
\begin{equation}\label{orel}
\frac{md^2}{\hbar}\omega_r=\frac{6}{\alpha^2}\sqrt{\frac{d}{w}}\sqrt{\frac{m\lambda}{\hbar^2 d}},
\end{equation}
in dimensionless form.
What we have done here is to use the number of bound states of the 
inverted potentials, which implies that we have in fact matched 
the number of bound states {\it excluded} by the presence of 
the repulsion. This is in similar spirit to the proposal of 
introducing an external oscillator discussed above. However, 
we avoid a dependence on properties beyond two-body physics.

In section \ref{multsect} we calculate critical values of 
$\omega_r$ beyond which configuration with several 
molecules in each layer become unstable and break into
smaller complexes.
The critical repulsive frequencies we find below are 
$md^2\omega_r/\hbar\sim 1-10$. Taking the specific case where 
$m\lambda/\hbar^2 d=20$, we find $md^2\omega_r/\hbar=7.78$. 
From (\ref{orel}) with $w=0.2d$ and $\alpha^2=2$, 
we find a larger value of $md^2\omega_r/\hbar=30$. 
This implies that we are in the unstable regime. However, 
we note that the parameter $\alpha$ is geometrical and 
somewhat arbitrary, constrained only by our desire to describe
the competition of repulsive and attractive terms in the
system. Our calculations below indicate that a better
value is $\alpha\sim 1.92$. This yields $md^2\omega_r/\hbar\sim 16$,
which is closer to the critical frequency. 

We are thus in
a situation where structures containing more than one
molecule in a single layer are most likely unstable in
realistic experiments. This is consistent with the 
conclusion of Ref.~\cite{artem2011c}
about the instability small complexes. However, as noted above
the experiments have external confinement and the structures
obtained for multiple chains below the critical frequency 
in later section could therefore appear as resonances in 
the confined system. The energetics and the path to breakup
for these systems is therefore an interesting question nonetheless.
With a harmonic model we can study such questions for large
number of particles in an essentially exact manner, given the
necessary approximation on the interaction terms.

In making a distinction between $m\lambda/\hbar^2 d$ and
$U$ (from the interlayer potential (\ref{e30})), we also 
leave open the possibility of working with more than one
type of molecule or populating different internal rotational 
states of the molecules in different layers, both of which
can modify the dipolar moment and potentially bind the 
structures we study here by reducing the overall repulsion.
This could possibly be achieved by externally applied 
electromagnetic fields and lasers 
\cite{giovanazzi2002,micheli2007,gorshkov2008}.
However, in the present study we will always assume that 
$\lambda=U$ and use just $U$ as the dipolar strength parameter.
In light of the extended discussion above, we
apply repulsive intralayer interaction parametrically
through the inverted oscillator to study the effects of 
inter-chain interaction for the purpose of the present work.

The oscillator potential is now defined for two molecules either in
different layers or the same layer. If a confinement is needed it is
straightforward to add and include a one-body
harmonic oscillator potential.  Without a one-body external potential
the relative motion separates from the free motion of the center of
mass which therefore becomes uninteresting. We shall therefore
consider only the relative motion.

\section{One molecule per layer}\label{oneptclsect}
The simplest many-body structure is found when only one molecule is
placed in each layer. This means that all the two-body interactions
are attractive and given in equation (\ref{e30}), and the repulsion in
equation (\ref{e60}) is not present.  Any pair of these molecules would then
form a bound state as shown previously \cite{armstrong2011,artem2011-1,artem2011-2}.
When the layers are far apart the potential is very shallow and the
attractive part extends to a distance, $r=nd\sqrt{2}$, proportional to
the distance between the layers. The attraction decreases with the
third power of the layer distance, and the binding energy decreases
correspondingly.  Therefore the interactions dominate between the
nearest neighbours.

\begin{figure}\centering
\includegraphics[width=0.5\textwidth]{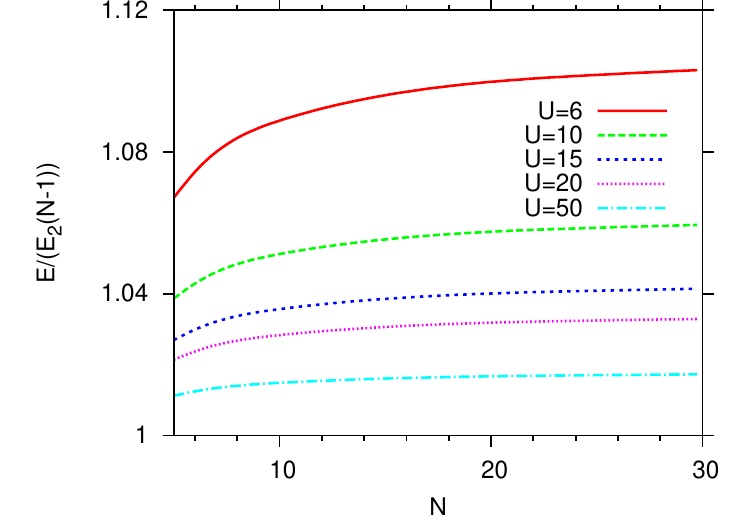}
\caption {Energies divided by two-body energy obtained with the 
potential in equation (\ref{e30}) times number of pairs of
nearest neighbours for
several different dipole strengths. These energies were obtained 
by restricting the potential to nearest neighbours only.}
\label{1stringEa}
\end{figure}

\subsection{Binding energies and radii}
The energies could be expected to be proportional to the two-body
energy, $B_2$, and linear in the number of interacting
neighbours pairs, $N-1$.  This was conjectured and estimated for large $N$ in
\cite{armstrong2011}, i.e.
\begin{equation} \label{e70}
\frac{B_N}{(N-1)B_2} \approx  \sum_{1}^{\infty} \frac{1}{n^2}= \frac{\pi^2}{6}\approx 1.645 \;.
\end{equation}
However, the assumptions are then that the interactions decrease as
$1/n^2$, and that no additional correlations arise from the string of
molecules. A better estimate should then be 
\begin{equation} \label{e80}
 \frac{B_N}{(N-1) B_2} \approx  \sum_{1}^{\infty} \frac{1}{n^3}= 
 \zeta(3) \approx 1.202 \;,
\end{equation}
where the first two powers are from the $d^2$-dependence and one power
is from the reduction of interaction strength, see equation (\ref{e50}). The
latter is an upper limit since the energy curve in figure \ref{endip} is
concave.  With only nearest neighbour interactions we get the energies
shown in figure \ref{1stringEa} for different dipole strengths. The
variation is largest for small $N$ while the energy per interacting
pair is leveling out for large $N$.  All curves are between $1$ and
$1.12$ with a monotonous decrease with interaction strength.  Thus
correlations always contribute on less than the $10\%$ level, and
increase as the system becomes spatially more extended and dilute.

\begin{figure}\centering
\includegraphics[width=0.5\textwidth]{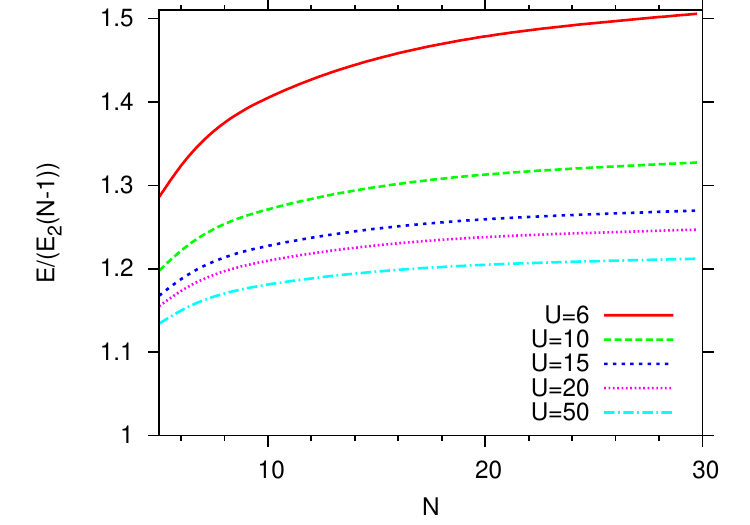}
\caption {Energies divided by two-body energy obtained with the potential in equation (\ref{e30}) times number of pairs of
nearest neighbours  for
several different dipole strengths. }
\label{1stringEb}
\end{figure}   

Inclusion of all interactions give the energies shown in
figure \ref{1stringEb}. The trends are the same with increase as
function of $N$ and decrease as function of interaction strength.
However, the actual values are now all between $1$ and $1.6$. The
large-$N$ asymptotics seems all to be above $1.202$ indicating
additional contributions from correlations which seems to approach
zero in the limit of strong interactions. 

\begin{figure}\centering
\includegraphics[width=0.5\textwidth]{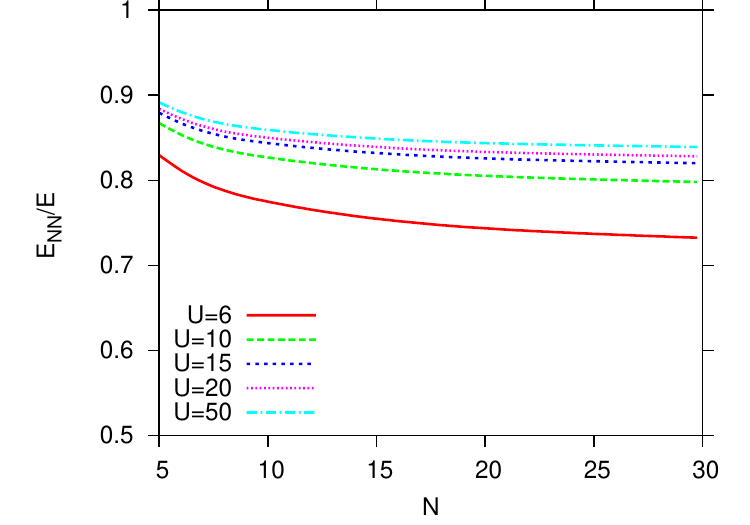}
\caption{  Energies in Fig \ref{1stringEa} divided by the energies in figure \ref{1stringEb}, 
which is the fraction of the total energy of the $N$-molecule string
  captured by the interaction of
  nearest neighbours as a function of chain length for several
  different dipole strengths.}
\label{NNE}
\end{figure}

The 20-50\% difference between energies in Figs. \ref{1stringEa} and
\ref{1stringEb} is due to nearest neighbour and inclusion of all
contributions to the energy.  The corresponding fraction is directly
shown in figure \ref{NNE} where the trend is decrease of nearest
neighbour contribution with $N$ and increase with interaction strength.
Again the strongest variation is for small $N$ while asymptotic
constant ratios are approached for large $N$. All ratios are between
$0.7$ and $0.9$, showing how good an approximation only nearest
neighbours would be, depending on strength and molecule number.  The
total number of interaction terms, $(N-1)^2/2$, increase compared to
the number, $N-1$, of nearest neighbour terms. However, the rather
strong fall off of the interaction with distance leads to the
relatively small values in figure \ref{NNE}. Even when a chain gets
longer, only the very nearest links in the chain are affected, and
constant ratios are approached.

The approximately linear dependence of energy on $N$ implies that the
separation energy, $B_{N-1} - B_{N} \propto B_2$, where the
proportionality constant crudely can be approximated by $\zeta(3) =
1.202$ (equation (\ref{e80})) for large $N$ and not too weak interactions.  The variation
with $N$ of the separation energies is around $0.2$ for all strengths.
The correlations therefore change smoothly with the number of molecules.

\begin{figure}\centering
\includegraphics[width=0.5\textwidth]{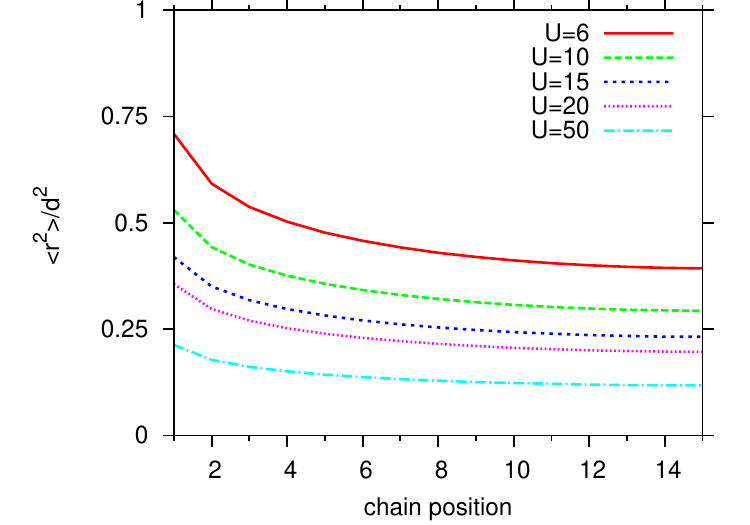}
\caption{Radius-squared as a function of the position in the chain of
  30 molecules.  Position one is the outermost position of the
  chain, and the radii are symmetric with respect to the center of the
  chain, so showing only the first half of the chain is necessary.}
\label{rad30}
\end{figure}

The wave functions for each of the states are also available. They are
essentially products of Gaussians of linear combinations of molecule
coordinates.  To get an idea of the structure,  we computed mean square
radii of the individual molecules measured from the common (projected)
center of mass of all molecules.  These radii are shown in
figure \ref{rad30} for a system of 30 molecules distributed in 30 layers
as function of the number of layers. The molecules in the outer layers
have largest radii intuitively correlated with the fewer efficiently
interacting neighbours.  Towards the central layer, the radii decrease
and become almost independent of layer position.  The radii decrease
with the strength of the interaction in agreement with the smaller
binding energies. This is consistent with the variational study 
in Ref.~\cite{wang2006}.

The absolute sizes are on average between $0.3d$ and $0.9d$ for the
strengths in figure \ref{rad30}. Clearly $d$ is the natural unit since
that is the extension of the potential within each of the layers. When
a radius is smaller than $d$ it means that the molecule mostly is
located inside the attractive well from the nearest neighbour.

\begin{figure}\centering
\includegraphics[width=0.5\textwidth]{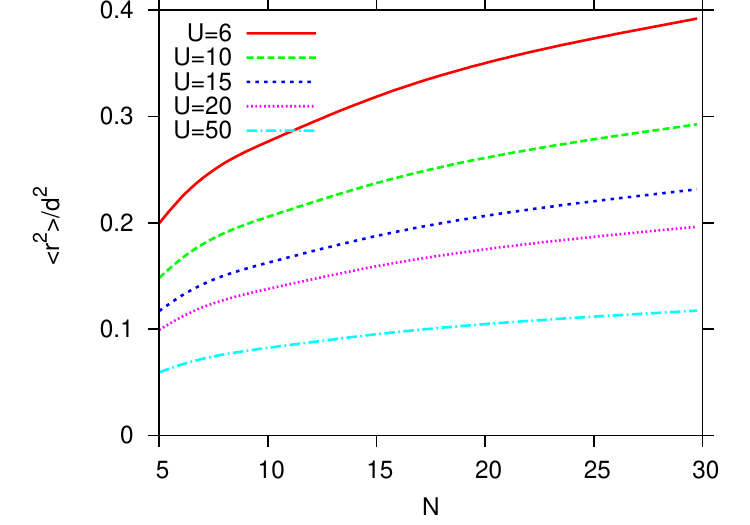}
\caption{Radius-squared of the central molecule in a chain of length
  $N$ as a function of chain length for several different dipole
  strengths.}
\label{radnum}
\end{figure}

To see how the radii evolve with molecule number, we show the central
radius as a function of molecule number in figure \ref{radnum}.  For
the same interaction strength, the size increases with molecule number.
  The outermost molecules are always the
largest, as they only have the influence of one nearest neighbour.  The
size of the outermost molecule also grows slightly faster than the
central molecules.

\begin{figure}\centering
\includegraphics[width=0.5\textwidth]{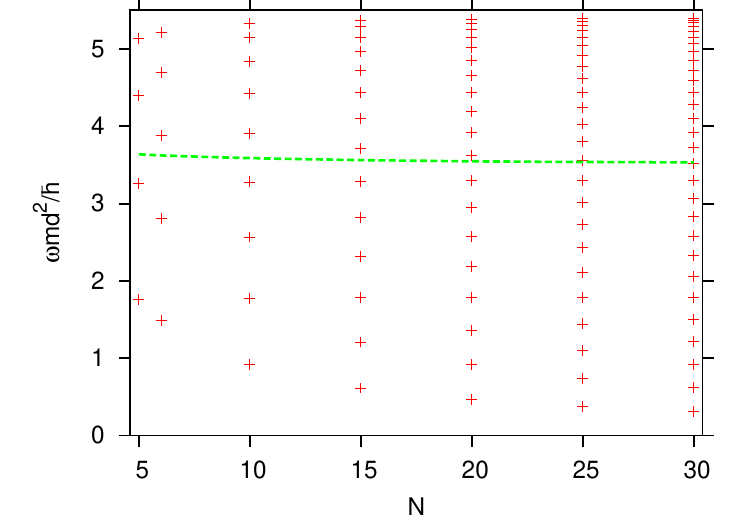}
\caption{Normal mode frequencies (points) as a function of molecule
  number for a dipole strength $U = 6$.  The line
  is the average value of the frequency at a given number of
  molecules.}
\label{freqN}
\end{figure}

\begin{figure}\centering
\includegraphics[width=0.5\textwidth]{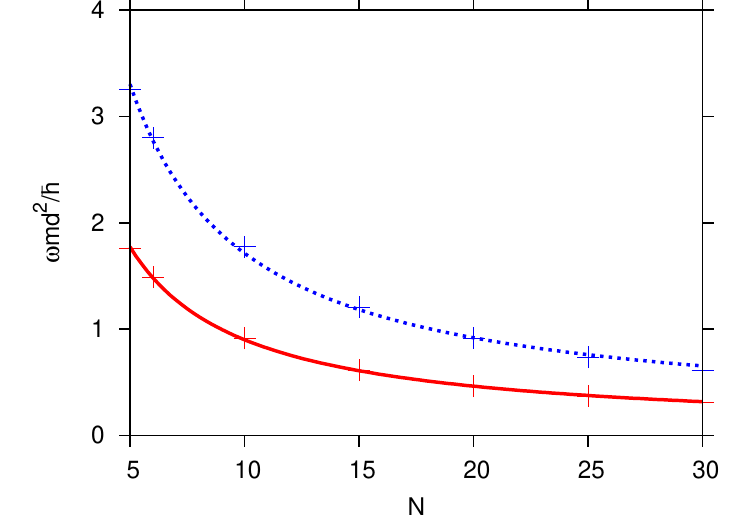}
\caption{Normal mode frequencies (points) as a function of number of molecules $N$ along with their fits to an $A/N+B$ curve, where $A$ and $B$ are fitting parameters for a dipole strength $U = 6$.}
\label{acoustic}
\end{figure}

\subsection{Normal excitation modes}
The harmonic oscillator solutions provide a series of frequencies
after the total separation of coordinates. They describe which modes
are easiest to excite. For $N$ molecules we always find $N-1$
frequencies related to relative motion (some of which
can be degenerate), and
one degree of freedom corresponding to the trivial free center of mass
motion. Note that we consider the limit where the layer is in the 
strict 2D limit, i.e. zero width. This means that all modes are transverse.
In a real experiment with a finite width, longitudinal modes are also 
possible but usually have higher excitation frequencies (in the 1D case
this has been discussed in Ref. \cite{astra2009}).

In figure \ref{freqN}, we demonstrate how the normal modes change with
chain length for a given interaction strength.  As the number of
molecules is increased, additional frequencies appear.  When adding
one molecule to a system of $N$ molecules, the $N-1$ frequencies from
this $N$-molecule system all decrease in magnitude. The new additional
frequency comes in above all the other frequencies, in such a way that
the average frequency decreases overall. The average frequency
multiplied by $N-1$ is in fact the ground state energy of the $N$-body
system, except for the correction from the shift in equation (\ref{e40}),
which to be precise should be multiplied by the number of interacting
pairs.  This is indicated by the horizontal line in figure \ref{freqN}.

\begin{figure}\centering
\includegraphics[width=0.5\textwidth]{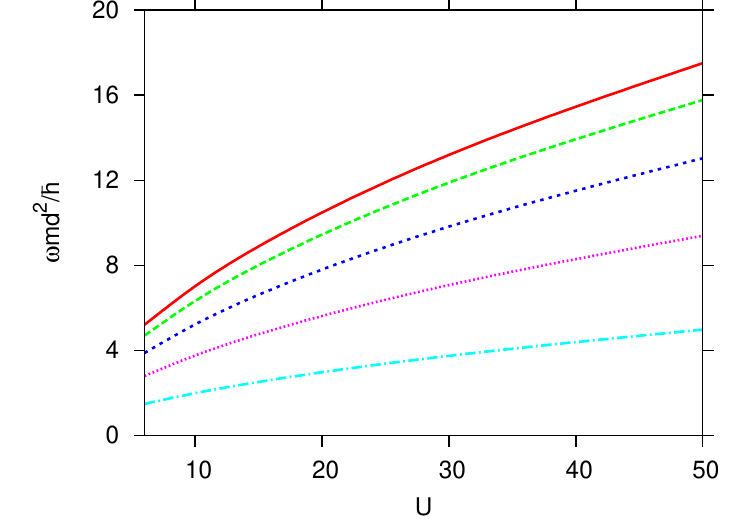}
\caption{Frequencies of the six-molecule system as a function of
  dipole strength.  The different lines connect frequencies of the
  same normal mode at different dipole strengths.}
\label{freq6}
\end{figure}

The normal modes in Fig~\ref{freqN} are similar to the acoustic branch modes
in a crystal with a single atom in the basis. In typical textbook presentations of
acoustic phonons, nearest neighbour interactions are taken into account and the 
spectrum has the form $\omega(k)^2=\tfrac{4K}{m}\sin(\tfrac{kd}{2})$, where
$K$ is the harmonic force constant of Hooke's law. Taking the naive harmonic approximation to the 
dipolar potential (obtained by expanding equation (\ref{e30}) to second order in 
$r/d$), we obtain $K=12D^2/d^5$. The speed of sound then becomes $c_0=\sqrt{\tfrac{12D^2}{md^3}}$
as discussed in Ref.~\cite{wang2008a}. By fitting the lowest modes of the chain as function 
of the chain length, we can get an estimate of the speed of sound in the harmonic framework 
used here. 

The possible wave vector in a string of length $Nd$ is $k=n\pi/Nd$, where $n$ is 
an integer $n>0$. We can now use the fit in figure \ref{acoustic} to obtain 
the speed of sound and we find $c=c_0/3.05$, a reduction by a factor of around 
three. The first obvious reason for the difference is that we have taken not 
only nearest neighbour interactions but all interactions into account in our 
model. However, as we saw in relation to the energetics, the terms beyond 
nearest neighbour play a minor role which was also noted in Ref.~\cite{wang2008a}.
The main difference is that the effective oscillator potential used here 
allows molecules to spread further in space, it is in a sense more shallow
than the naive approximation around $r/d\ll 1$. This results in a less rigid
string with smaller tension and therefore a reduction in the speed of sound. However, 
from the energetics alone it seems clear that this is a much better approximation, 
and we have thus shown that also the single-chain modes should be re-evaluated.
Notice also that in the naive approach, $\omega(k)^2$ scales with $U$, whereas
we find a scaling that is softer ($U^{\alpha}$ with $0<\alpha<1$) due to the 
reduced oscillator frequency. 

The tension of the chains/strings is directly related to the average radius $\langle r^2\rangle$ 
of the particular potential used since this indicates the strength of the interaction 
and how fast the chain returns to its equilibrium position after a disturbance. As 
shown in Ref.~\cite{zinner2011a}, the real potential has $\langle r^2 \rangle$ 
slightly smaller than the harmonic approximation used here. The naive harmonic
approach has $\langle r^2\rangle$ that is an order of magnitude smaller than 
both the real potential and the effective harmonic potential used here. 
This is another indication that the spectrum produced here is a significant 
improvement over the naive approach. Based on the fact that the real dipole 
potential has slightly lower $\langle r^2\rangle$, the true speed of sound 
should also be somewhat higher than our estimate.

In figure \ref{freq6} we show how the dipole strength affects the
normal mode frequencies.  They all increase, as an increase of the dipole
strength essentially makes the chain stiffer. On the other hand, the
curves in figure \ref{freq6} are not parallel, and the frequencies
diverge away from each other. The higher the frequency of the mode,
the more it grows with increasing dipole strength. This is different from 
the usual model of acoustic phonons with only nearest neighbours, and is 
due to the inclusion of next-nearest neighbours and so on. The latter interactions
tend to suppress short wave length modes.

\begin{figure}\centering
\includegraphics[width=0.5\textwidth]{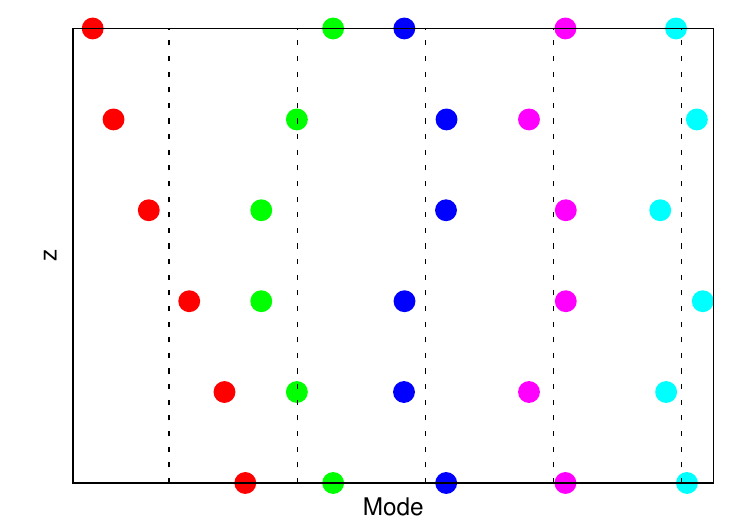}
\caption{Normal modes for a chain with six molecules and $U =10 $.  The vertical axis shows the molecule's layer
  position.  In the horizontal direction, the modes are arranged in
  increasing order of frequency (not to scale).  The dashed vertical
  line shows the equilibrium position of the molecules, which are
  shown as points.  The frequencies of the modes are, from left to
  right, 2.00, 3.77, 5.23, 6.34, and 7.03 in units of $\hbar/(m d^2)$.}
\label{modes1str}
\end{figure}

The structure of the normal mode excitations are displayed in
figure \ref{modes1str} for a chain of six molecules. We show the maximum
displacement amplitudes in one direction of the 2D structures.  The
lowest energy mode has one node in the vibration of the string, the next
mode has two nodes, and so forth.  This ordering remains for all
dipole strengths, and the amplitudes of the molecules change very
little.

It should be noted that this pictorial representation is in one
dimension. The other direction is degenerate but independent, which
implies that linear combinations of each of these one-dimensional
modes are equally appropriate.  The displacements could happen along
any two independent directions in the $xy$-plane. This is reflected in
the fact that we always find two degenerate modes for each normal 
mode frequency, both for single chains and for several chains per 
layer as discussed below.
For example, we
could choose a radial and tangential mode in the planes instead of the
Cartesian one-dimensional modes illustrated in figure \ref{modes1str}.
Still they would have to present nodes at the same layers, where the
molecules remain at their equilibrium position.

\section{Multiple chains and layers}\label{multsect}
Molecules in different layers still attract each other as parametrized
in equation (\ref{HOpot}). However, more than one molecule in the same layer
immediately introduces the repulsion between these molecules.  We
parametrize this additional interaction in equation (\ref{e60}) where we
treat the repulsive frequency, $\omega_r$, as a free parameter.  All
structure calculations are independent of the constant shift which
only enters in energy calculations.  We therefore first discuss
structure variation as $\omega_r$ is increased from zero.  Eventually
there is a point when the repulsion is too strong compared to the
inter-layer attraction, and the system is no longer bound.  This is
seen as at least one imaginary normal mode frequency solution to the
full $N$-body problem.

\begin{figure}\centering
\includegraphics[width=0.5\textwidth]{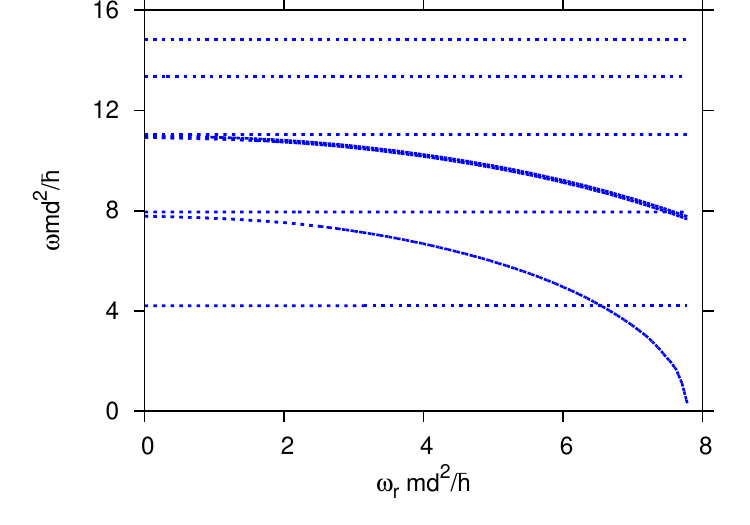}
\caption{ Evolution of the normal mode frequencies as a function of
  the intra-layer repulsive frequency, $\omega_r$.  The modes are for
  two strings of six molecules with $U=20$. }
\label{freqRep}
\end{figure}

\subsection{Energies and repulsion}
We now solve for $N$ layers and two dipolar molecules in each
layer. The resulting frequencies are shown in figure \ref{freqRep} for
$N=6$ as function of the repulsive frequency.  There are $11$
normal mode frequencies for these $12$ molecules, one mode is
related to the free center of mass motion.  Five of these modes do not
depend on the repulsive frequency.  They correspond to the modes of the
singly occupied layers (similar to those seen in figure
\ref{modes1str}), but with the frequencies increased by a factor of
$\sqrt{2}$. This reflects that the single string degrees of freedom is
repeated by the pairs in each layer.  The strings move in phase as if
they were alone, and are like the modes shown in figure  \ref{modes1str}
with pairs following each other instead of single molecules.  The
factor, $\sqrt{2}$, in the energy arises since the total energy is the
square root of the sum of energies \cite{armstrong2010}. Equivalently, for
two chains the number of attractive pairs double and in turn the 
string tension doubles. The factor comes from the acoustic mode frequency
which depends on the square root of the tension. For $M$ chains, the
enhancement is then $\sqrt{M}$. While this seems at odds with a 
classical picture of a collection of chains in 2D layers, the 
molecules and chains in our model are not localized and therefore
adding a chain enhances the chain in the same way, irrespective of the 
number of chains one starts from.

The remaining normal modes in figure \ref{freqRep} appear as three
doubly degenerate modes decreasing with the repulsion. They represent
relative motion where the two single string degrees of freedom are
mixed.  The degeneracy is due to the reflection symmetry in a central
plane parallel to the layers.  Two of these three degenerate
frequencies coincide resulting in quadruple degeneracy. These modes
correspond to motion of the molecules in the four middle layers.  The
last degenerate frequency in figure \ref{freqRep} decreases to become
the lowest mode, and eventually reach zero for a critical repulsive
frequency $\omega_c$. The system afterwards is unstable corresponding
to imaginary solutions of the energy.

\begin{figure}\centering
\includegraphics[width=0.5\textwidth]{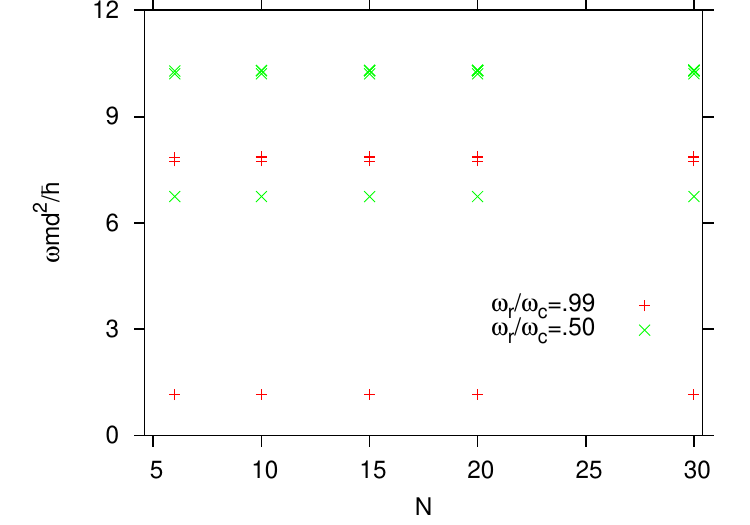}
\caption{Normal mode frequencies (points) as a function of number of layers
  for two strings with a dipole strength $U = 20$ and two different repulsive strengths.  
  The modes shown here are only the intra layer modes.  The cohearant 
  full string modes behave in the same way as shown in figure \ref{freqN}.  }
\label{freq2Nstrngs}
\end{figure}

The behaviour of the frequencies with increasing the number of layers
is seen in figure \ref{freq2Nstrngs} (similar to figure \ref{freqN}).  
Unlike the single string cases, we see that they are almost completely 
flat. This is analagous to the optical frequency modes in crystals
with more than one atom at each lattice site, and we may refer to
these as optical (normal) modes.
As the number of layers is increased, more and more frequencies 
appear in the upper branch, which are not resolved on this figure's 
vertical scale.  They are nearly degenerate in any case since they 
are modes of the central layers.  The effect of repulsion is to push 
the frequencies of the outermost layers to zero, which was also 
seen in figure \ref{freqRep}.

\begin{figure}\centering
\includegraphics[width=0.5\textwidth]{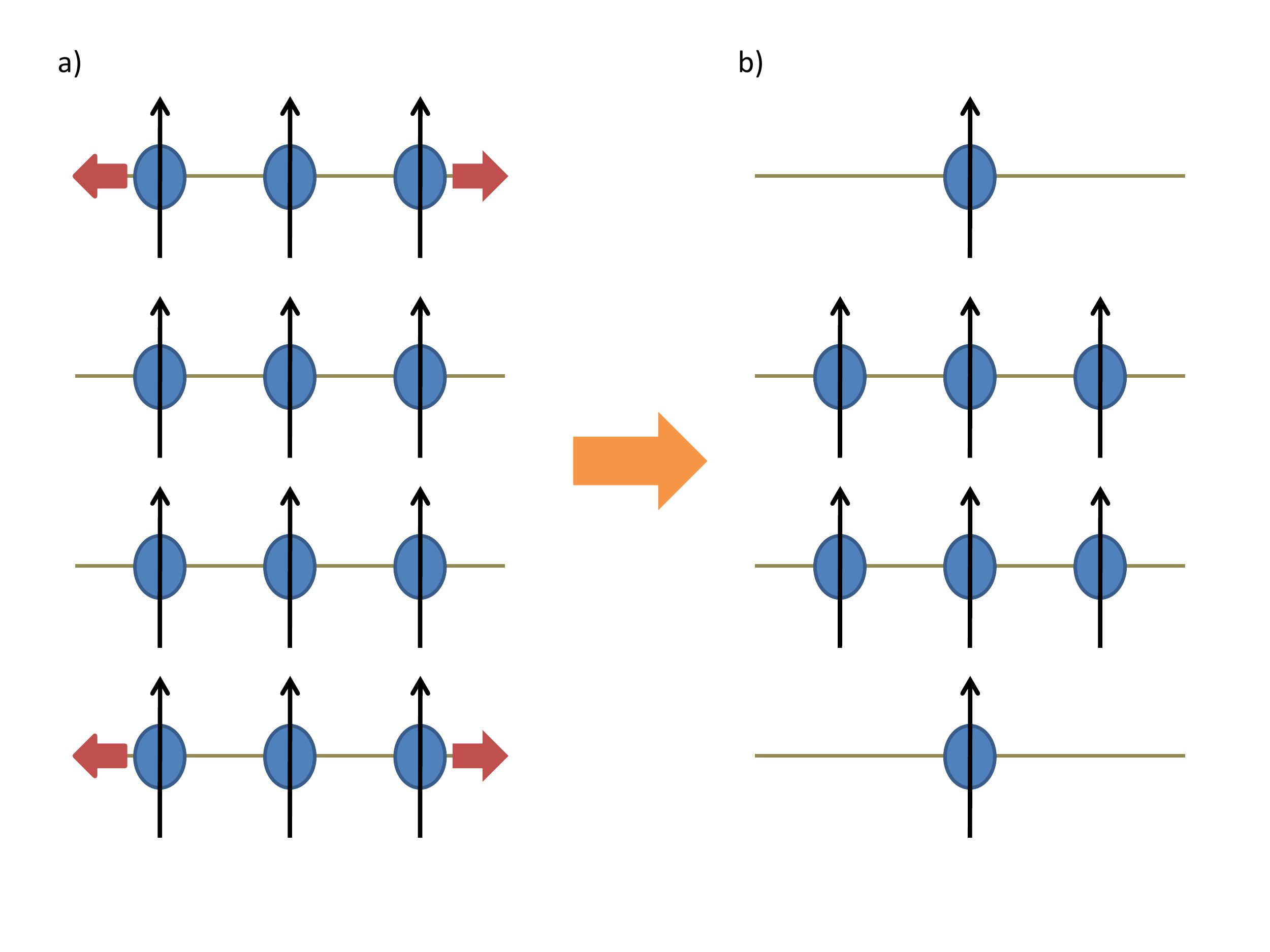}
\caption{Schematic illustration of the unstable mode for three chains with four particles in each chain. The 
picture shows a 1D setup for clarity (the 2D case has a similar cylindrical structure). a) The instability of the
outermost molecules in the top and bottom layers. b) The final state of the system after dissociating the two outer
molcules.}
\label{break}
\end{figure}

The structures of the two optical modes corresponding to the 
degenerate energy that approaches zero are relative motions between the two molecules in
each of the outermost layers.  The instability is then that the
amplitudes of the vibrational motion between the pairs of outer
molecules become too large to return to equilibrium.  The degeneracy
means that all four molecules then separate simultaneously as illustrated in 
figure \ref{break}.  Then the
whole system immediately falls apart, as all shorter chains break at
smaller repulsive frequencies. 

If we increase the number of layers the results would be very similar.
Only two differences appear. The first is that the number of
repulsion-independent (horizontal) solutions increase, because the
number of single chain solutions increase.  The second difference is
that the degeneracy increases for the highest of the optical normal
modes, which involve molecules in the intermediate layers. They are
essentially independent of the length of the chains, see figure \ref{freq2Nstrngs}.  
The lowest
frequency optical mode which eventually approaches zero 
is also almost independent of
chain length, and the unstable mode is again related to the outermost
molecules.

If we instead increase the number of chains, the frequencies of the
resulting solutions would appear almost precisely as in
figure \ref{freqRep}.  Again only two differences appear. The first is that
now the repulsive-independent (horizontal) solutions are obtained
from single chain frequencies by multiplying by the square root of the
number of chains.  The reason is again that energies are obtained as
square roots of average frequencies as explained above.  The second
difference is that the degeneracies increase for the two decreasing
optical mode solutions. The highest of these correspond to normal modes involving
molecules in the intermediate layers. They are essentially
independent of the number of chains since they move pairwise without
influence from the other layers.  For each
additional chain the degeneracy of the terminal layers increases by two, that is one
corresponding to each end-layer structure.  The unstable modes are now
more complicated and at least one of them always involves at least one
molecule of small amplitude. This implies that at least one molecule
remains after the amplification of the amplitudes of the unstable
mode and a subsequent breakup of the complex.

\subsection{Critical stability}
The structure close to the instability can be seen directly from the
behaviour of $\langle (r_i-R)^2\rangle$ for the individual molecules as
function of the repulsive frequency.  Here $R$ is the center of mass
coordinate projected on a plane.  This quantity can be obtained
since each molecule is treated as distinguishable.  Still, identical
bosons in the same layer always have the same radii.  As $\omega_r$
increases all radii of interior molecules increase marginally whereas
the terminal positions increase substantially faster.  However, when
$\omega_r$ is very close to the critical frequency, $\omega_c$, the
internal molecule-radii perhaps double their size but the terminal
sizes meanwhile increase by an order of magnitude and eventually
diverge. The reflection symmetry in a central plane can be maintained
throughout.  Due to the circular symmetry, the system can be described
as a cylinder, whose radius changes down its length, becoming wider
towards the ends.

The critical frequency, $\omega_c$, can be found numerically by
solving the Schr\"{o}dinger equation. For $N$ layers and the same
number of molecules in each layer, we label the molecules from $1$ to
$N$ along the first chain and continue the consecutive labeling chain
after chain.
A formula for the critical frequency, $\omega_c$, and the 
shift, $\alpha$, can be obtained analytically 
by equating the repulsive and attractive potential energies as seen by 
one of the molecules in the outermost layer where the instability takes place
\begin{eqnarray}
\frac{1}{2}\mu_{ij}\omega_c^2(x_{ij}^2+y_{ij}^2)(M-1)+
\frac{1}{2}\mu_{ij}\omega_c^2\alpha^2d^2(M-1)=&& \nonumber\\
\frac{1}{2}\mu_{ij}(M-1)\sum_{j=2}^N\omega_{1j}^2(x_{ij}^2+y_{ij}^2)
+\mu_{ij}(M-1)d^2\sum_{j=2}^N\omega_{1j}^2(j-1)^2&&,
\end{eqnarray}
where $\omega_{1j}$ are the input frequencies reflecting the interaction 
between a particle in the outermost layer and layer $j$.  Eliminating 
common factors allows one to obtain an expression for the critical 
frequency and the shift factor
\begin{equation}\label{e90}
\omega_c^2=\sum_{j=2}^N\omega_{1j}^2,
\end{equation}
and
\begin{equation}
\alpha^2=2\frac{\sum_{j=2}^N\omega_{1j}^2(j-1)^2}{\sum_{j=2}^N\omega_{1j}^2},
\end{equation}
where the frequencies on the right hand side of equation (\ref{e90}) are the 
input frequencies determined in equation (\ref{e40}). 
These expressions hold for a system of equal masses and would have to 
be changed for unequal masses.
The
result provides the value for $\omega_{c}(N)$ which explicitly is
labeled with the dependence on layers.  This is a remarkably simple
and very general result independent of the number of chains.
The expression for $\alpha$ depends only very weakly
on $U$ and $M$, and we find $\alpha^2 \sim 2.26$ numerically 
for most values of $N, M,$ and $U$.  
The expression for $\omega_c$ exactly reproduces the numerically determined 
values. This is completely independent of $\alpha$ since, as we will discuss 
below, $\alpha$ is determined by equating energies of different 
configurations, i.e. it is a pure
shift of the energy which will not influence the stability of the system. 
Whereas for $\alpha_c$ this is just a good initial guess at the proper 
value, which will be commented on in following sections.

Thus, $\omega_c(N)$ increases with number of layers, but only slowly,
since smaller and smaller frequencies from more distant molecules are
added to the sum in equation (\ref{e90}).  The hierarchy of stability is now
easily established, because the sequence follows the size of
$\omega_c(N)$.  If a given number of layers becomes unstable, the
structure with one layer less is also unstable. On the other hand,
we find that the critical repulsion is independent of the number
of chains.

So far we considered systems with the same number of molecules in all
layers.  The normal modes are then related to symmetric structures
essentially localized as motion within planes and relative motion
between groups of molecules in different planes. A different system
could have fewer molecules in the outer layers than the series of
identical layers in the interior, an example is shown in figure \ref{break}b). 
The hierarchy of stability follows
from the number of repulsions in the outer layers, that means one molecule
is most stable, two second most, etc.

The critical frequency for a configuration with one molecule in the
outer layers, $\tilde\omega_c$, is found to be
\begin{equation} \label{e100}
  {\tilde\omega}_{c}^{2} = \omega_{c}^2(N-2) + \frac{1}{M} (\omega_{12}^2 +
  \omega_{2N}^2)  \;,
\end{equation}
where we used the definition in equation (\ref{e90}), the labeling are that
molecules $1$ and $N$ are the lone outer molecules, and molecule $2$
is in the second layer with $M$ molecules.  Interestingly, the ratio,
${\tilde\omega}_{c}^{2}/ \omega_{c}^2(N)$, seems to be independent of the dipole
strength.

\begin{figure}\centering
\includegraphics[width=0.5\textwidth]{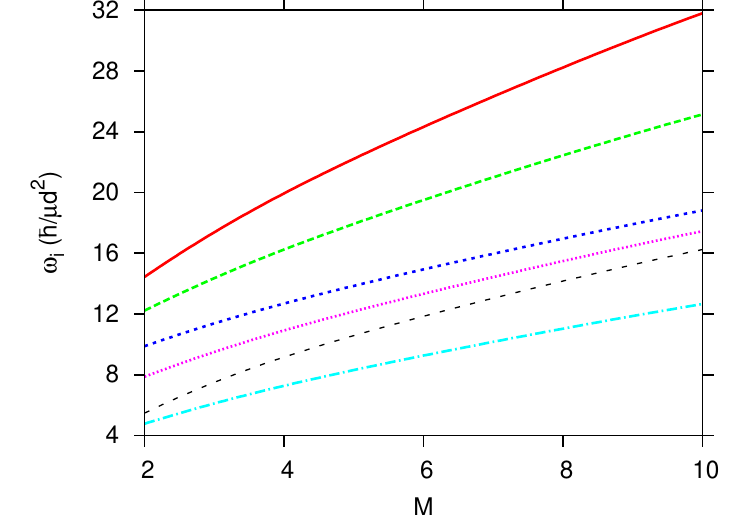}
\caption{ Single chain modes as a function of number of chains, $M$,
  for a broken system of 6 layers (where now there is only one
  molecule in the outermost layers).  The lines connect the same
  normal mode at a given number of chains.  The dashed black line is
  the frequency of the degenerate mode which is not the breaking mode
  frequency taken when $\omega_r\approx\omega_c$.  The dipole strength
  was $U=20$.}
\label{freqbrk}
\end{figure}

The behaviour of the normal modes and normal mode frequencies is
similar to the complete system.  The total number of remaining modes
is $M(N-2)+1$.  The $N-1$ single chain modes remain, and they appear
qualitatively just as seen in figure \ref{modes1str}.  These frequencies
increase with the number of chains, and are shown in
figure \ref{freqbrk}, but no longer in the square root dependence that
was seen in the complete system.  As with the unbroken system, the
single chain frequencies do not change with the repulsive frequency.
The remaining modes fall into two degenerate groups of equal size,
$2(M-1)$.  One group approaches zero as the repulsive frequency is
increased, while the remaining modes are at a frequency between the
two lowest single chain modes.  The group approaching zero are modes 
involving the outermost full layers, which become unstable as the 
frequency is increased.  The other group are modes in the central layers.

Another question is which structure the unstable mode has when only
one molecule occupies the outer layers. The unstable mode is moved
to the next-to-outermost layers where the molecules close to the
critical frequency then behave as in the system where they were in the
outer layers.  The resulting configuration is then most likely one
molecule in each of the two outer layers in each end.  This
configuration is stable because the decrease of $\omega_{c}^2(N-2)$ to
the value $\omega_{c}^2(N-4)$ is compensated by the added dominating
term, $\omega_{12}^2$, as seen in equation (\ref{e100}).  Increasing the
repulsion would repeat the process of instability of the next fully
occupied layers.  The system seems to prefer to approach single chain
structure. 

If we decrease the number of molecules in the outer layers by one
molecule at a time, the instability is seen as degenerate modes of
varying frequency approaching zero for sufficiently strong repulsion.
The degeneracy corresponds to an instability in the
outer layer until the outer layer has just more than half as many
molecules as the second layer. Then the degeneracy changes to
correspond to instability of the molecules in the second layer. This
reflects that the instabilty is initially related to the outer layer
until a number of molecules is removed from that layer and the second
layer contains the unstable mode.  Regarding the critical frequencies, 
for the outermost layer, the critical frequency is
\begin{equation}
  {\tilde\omega}_{c}^{2} =\omega_c^2(N-2)+\omega_{1N}^2
  +\omega_{2N}^2+\frac{h}{M-h}\sum_{k=2}^{N-1}\omega_{1k}^2,
\end{equation}
where $h$ is the number of holes in the outermost layer, and $\omega_{1j}$ 
is the interaction frequency between layer one and layer $j$.  For the 
molecules in the penultimate layer, the critical frequency is
\begin{equation}
  {\tilde\omega}_{c}^{2} =\omega_c^2(N-2)+\frac{M-h}{M}(\omega_{12}^2+\omega_{2N}^2).
\end{equation}
Whichever of these two frequencies is lower for a given configuration 
determines in which layer the instability lies.  

In summary, for the frequencies, there are a total of $NM-1$ normal mode
frequencies.  A subset, $N-1$ of them, are single string modes with
frequency equal to $\sqrt{M}$ the frequency of the corresponding
single string mode.  There are then many degenerate modes, $2(M-1)$ of
them go to zero as the repulsive frequency is increased.  This is
because the chain breaks by removing the repulsion in the outermost
layers, releasing all but one molecules.  The remaining degenerate
frequencies cluster into groups, and are often very close to each
other in magnitude.  They also decrease as the repulsive frequency is
increased, similar to what is seen in figure \ref{freqRep} for two
strings.

\begin{figure}\centering
\includegraphics[width=0.5\textwidth]{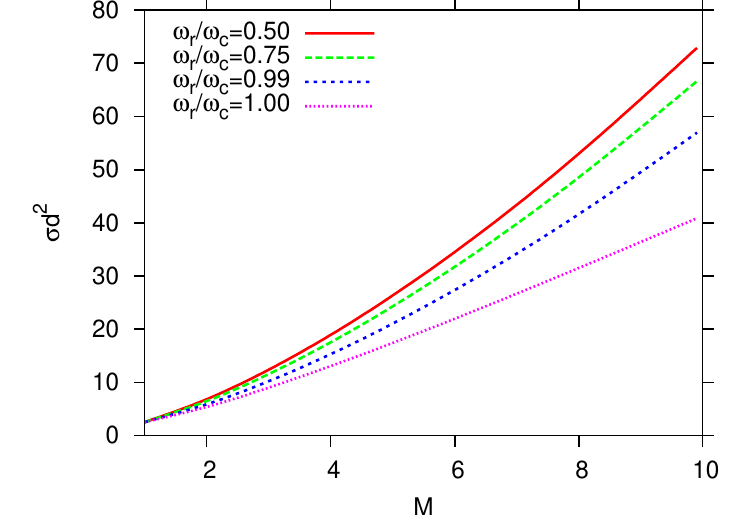}
\caption{Density of the central molecules as a function of number of
  chains.  Each chain is eight molecules long and the dipole strength
  was $U=20$.}
\label{density}
\end{figure}

\section{Density and energy relations} \label{denssect}
The repulsive frequency and the repulsive shift are so far left
as free parameters, although we did 
discuss at length how one can relate them to two-body properties
in section \ref{methodsect}.  
The shift is only
related to total energies without any influence on structures. We
need to define a zero point to measure from but this requires more
than two molecules, and the criterion easily depends on which
system was used for the calibration. In the next subsections we deal
with these two parameters in the in-layer repulsion.

\subsection{Density versus repulsive frequency}
In a system with many layers, the density
depends on the relative position of the layer. To be specific we consider
the central layers where the density
is rather constant across a few layers.  The outer layers
close to the critical stability are on the other hand expanding as the
breaking point is approached.  Already for a relatively small number
of layers this expansion only affects the outer layers.  We thus
define the density through the average radii of the central layers,
i.e.
\begin{equation} \label{e110}
 \rho = \frac{M}{\pi\langle \bar{r}^2\rangle} \;,\;\;\;
 \langle \bar r^2\rangle = \frac{1}{M} \sum_{k=1}^{M} \langle (r_k-R)^2 
 \rangle \;,
\end{equation}
where this definition is valid for non-identical molecules.  The
densities of the central layers defined in equation (\ref{e110}) are seen in
figure \ref{density} for different repulsions as functions of number of
molecules in each layer.  These curves are essentially independent of chain length, 
but increase with molecule number and decreasing repulsion.

\begin{figure}\centering
\includegraphics[width=0.5\textwidth]{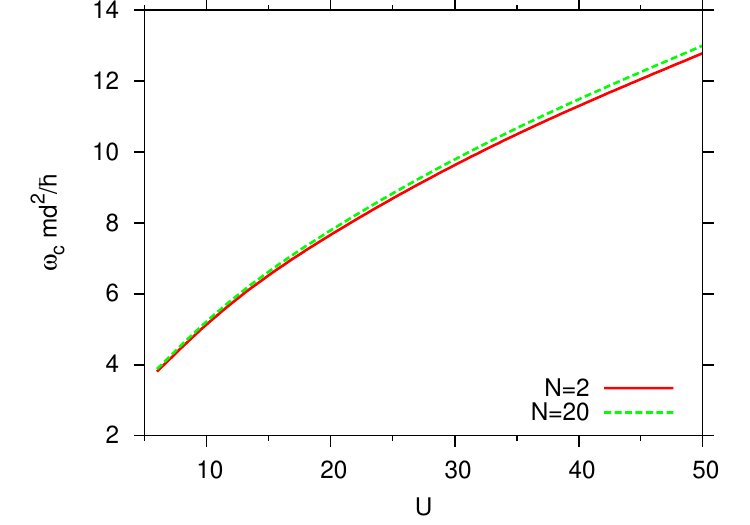}
\caption{ Critical frequency for two chains of differing lengths as a 
function of interaction strength.  }
\label{omegaD2}
\end{figure}

\begin{figure}\centering
\includegraphics[width=0.5\textwidth]{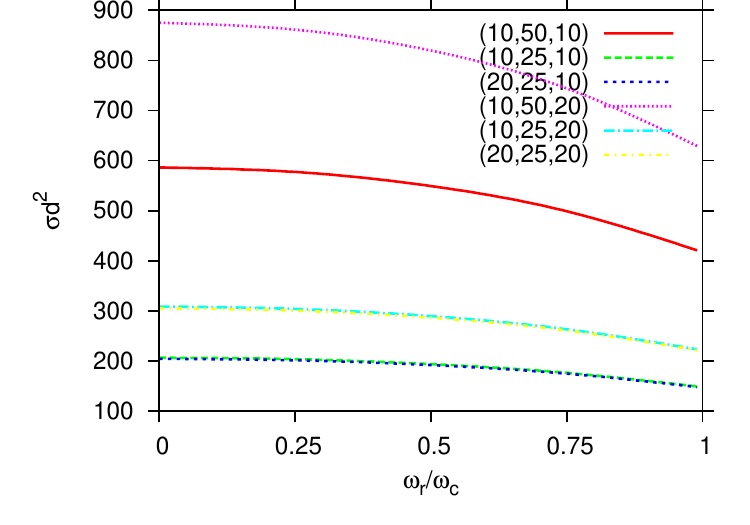}
\caption{ Densities as a function of the ratio of the repulsive frequency 
to the critical repulsive frequency.  This is plotted for several values of 
the parameters $N, M,$ and $U$, and the curves are identified by the label $(N,M,U)$.  }
\label{densities}
\end{figure}

The density of the central layers are essentially independent of chain length.  
It does depend, however, on the dipole strength, molecules per layer, $M$, and 
the repulsive frequency, i.e., $\rho\propto\sigma(M,\omega_r,U)$ with a very 
weak dependence on $N$.  The dependence on $M$ is a straightforward $M^{3/2}$, 
which comes from, in eq (\ref{e110}), with one power of $M$ in the numerator, 
and it is found that $\langle r_k^2\rangle$ goes like $M^{-1/2}$.  The critical 
frequency, and hence the critical density, below which the system becomes 
unstable, depends on the parameters as mentioned above.  We can state 
explicitly the dependence on $M$, and thus try to isolate its dependence 
on frequency, and thus $U$:
\begin{equation}
\rho_c\propto M^{3/2}\omega_c(N,U),
\label{densrelation}
\end{equation}
though the dependence on $N$ is very weak.  The dependence on $\omega_c$ 
is linear, and the dependence of $\omega_c(N,U)$ on $U$ is seen in figure \ref{omegaD2}.  
One can see that the dependence on the length of the chain, $N$, is indeed 
very weak.  However, in looking at densities of central layers, one must 
have a chain of 4-5 layers in order for a ''central layer'' to be clearly 
defined.  A plot of the density as a function of intermediate values of 
the repulsion is seen in figure \ref{densities}.  One can see that when the 
chain lengths are doubled, the curves  are almost indistinguishable.  The 
relations between the lines that differ in the other parameters are as 
described in eq (\ref{densrelation}).  The behaviour of the curve is not 
analytical, but the ratio $\rho(\omega_r=0)/\rho_c\doteq 1.39$, where 
$\rho(\omega_r=0)$ is the density with zero repulsion, is seen for all 
the sets of parameters. 

In section \ref{methodsect}, we discussed the option of relating 
the magnitude of the repulsive 
frequency to density (see (\ref{ocrit})), but did, however, caution 
that this is not a very desirable approach to effective interactions
in the $N$-body problem.
The density that we obtain within our model can become very large. In fact, 
if we insert the obtained densities of order $\sigma d^2\sim 10-100$ in (\ref{ocrit}), 
the frequency obtained is larger than the critical frequency by an 
order of magnitude already for $M=3$ and increases rapidly with $M$.
This shows the 
dominance of the attractive interlayer forces, i.e. 
the system contracts into
denser configuration than the typical average experimental density.
We stress again that (\ref{ocrit}) is a questionable relation but
note that the alternative way of fixing the repulsion through the 
bound state structure of the potentials also predicts a repulsion
slightly beyond the critical value. So in a system where the inter-
and intralayer dipolar strength is the same, we expect the 
configurations with multiple molecules in single layer to be 
unstable.
 
An important point here is that the critical frequency is essentially
independent of $M$, i.e., if we add another chain to the system for
a fixed value of $\omega_r<\omega_c(N,U)$. On the other hand, the density
goes up since the additional attractive interlayer terms that balance the 
extra repulsion in each layer will make the system contract.
According to (\ref{ocrit}) this implies that $\omega_r$ should be increased
and it will then quickly exceed $\omega_c(N,U)$. Again, it is 
unfortunate to change the two-body interactions in response to 
a property of the total system. By allowing the repulsion to be 
a free parameter we can study the detailed competition between 
repulsive and attractive term in the system which was one goal of the 
current study.

\subsection{Energy versus repulsive shift}

The relative energies depend on the repulsive shift in
equation (\ref{repshift}).  A consistent choice of scale factor, $\alpha$,
independent of chain $M$, and layer, $N$, numbers is not possible.
The problems arise, because instability towards some continuum
structure is found through the lack of solutions, and totally
independent of the energy of the configurations.  However, the
repulsive shift is precisely able to adjust energies such that the
wave function and energy criterion coincide for specific geometries.
The main effects can be included by an appropriate choice. 

The constant, $\alpha$, may be obtained by comparing the energies of
one system at the point of instability with the other system formed
after breaking apart.  The constant is adjusted so that the energies
of these systems are the same at the breaking frequency.  Let us
choose the simplest structure with three molecules in each layer and
compare to the system where two molecules are removed from each of the
end-layers.  This requirement leads to $\alpha\sim 1.92$, which makes the
repulsive shift somewhat weak compared to the attractive one at the 
breaking frequency.  This choice of $\alpha$ has the nice feature that 
as the number of molecules per layer is increased, the energy becomes 
more negative, but does not run away to negative infinity.  This is a 
nice feature since we know that the breaking frequency is independent 
of $M$.  However, this value of $\alpha$ will not  match energies when 
the system is driven to its breaking frequency.  A better form for the 
constant might be
\begin{equation}
\alpha^2=\alpha_0^2\pm\alpha_1^2/M,
\label{newalpha}
\end{equation}
where $\alpha_0$ is obtained by fitting to the quadratic dependence of 
the energies of $M$ in systems where $M \geq ~5$.  The second 
constant, $\alpha_1,$ is determined by matching the energies of a complete 
system, and the system where one molecule has been removed from both of 
the outermost layers (where it is assumed that the $\alpha_0$ term will 
cancel).  This combines two desirable features, controlling the energy 
dependence on $M$ and matching with the broken system.  These constants 
were determined for the system $(N,M)=(6,8)$ and $U=20$ to 
be $(\alpha_0,\alpha_1)=(1.9761,3.0123)$, and the lower sign is taken in 
equation (\ref{newalpha}).  These constant do what is prescribed of them, but 
do have drawbacks.  For example, for $M=2$, the overall sign is negative, 
which makes it an additional attractive shift.  Also, while the energies 
at breaking match at $M=8$, at $M=9$ they already are mismatched by 1.2\%.  
Clearly one value of $\alpha$ cannot accurately reproduce all the
desired properties of the total energies. A reasonable average choice
is $\alpha\sim 1.92 $ but more accuracy can be achieved by focussing
locally on a specific system and tune the value of $\alpha$. 

This value
of $\alpha\sim 1.92$ was cited in the discussion of the two-body repulsive interactions
in section \ref{methodsect} and leads to a reasonably low $\omega_r$, which 
is, however, still larger than the critical frequency for breakup. 
Again we remark that $\alpha$ is a geometric quantity related to the 
form of the inverted oscillator that provides the repulsion. 

By increasing the number of chains, or increasing the number of
molecules in each layer of the cylinder, we are essentially adding
another dimension to the system, a number of molecules per layer in
addition to the length of the chains.  For a system of $M$ chains of
$N$ molecules, one can count the number of attractive and repulsive
pairs in the system.  If we keep to the nearest neighbour level of
attraction, then the number of attractive pairs is $M^2(N-1)$, and
the number of repulsive terms is $M(M-1)N/2$.  Their ratio is then
\begin{equation}
\textrm{Attractive pairs}/\textrm{Repulsive pairs}=\frac{2M(N-1)}{N(M-1)}.
\end{equation}
This ratio appears, for example, in the bilayer system ($N=2$), where if one keeps
adding more pairs of molecules (one in each layer per pair), 
the ratio of attraction to repulsion
decreases from two ($M=2$) to one ($M\gg 1$).  One might expect that this would
influence the critical repulsive frequency, but this frequency does
not change with the addition of more chains.  In the nearest neighbour
approximation, it would also not change with increasing the chain
length.  It is the interaction beyond the nearest neighbour that can
affect the critical frequency.  The relation can, however, affect the
energetics of the system, depending on the relative strengths of the
attraction and repulsion.  As mentioned before, this indicates that a 
scale factor defined for a system with only three chains $(M=3)$, would 
be too large if used for systems with large $M$, as the ratio of 
attractions to repulsions falls as $M$ increases (for fixed $N$).

\begin{figure}\centering
\includegraphics[width=0.5\textwidth]{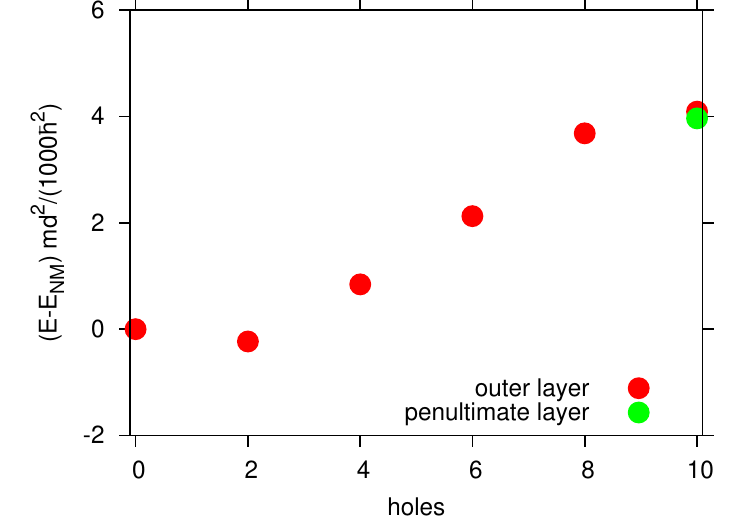}
\caption{ Energies relative to the full six layers of ten molecule 
system, i.e., $(N,M)=(6,10)$ as molecules are removed from the 
outermost layers.  molecules are removed one-by-one from both outer 
layers, in order to preserve the symmetry about the center layer 
of the system.  When performing the calculation in a system where 
molecules have been removed, the repulsive frequency used is the 
critical frequency of the parent system.  This was not done for 
the the system where ten molecules where removed, since the 
$\omega_c$(8 holes)$>\omega_c$(10 holes).   For this system with 
ten holes, the calculations were done for the two different geometries:  
five holes in each outer layer, and with four holes in the outer layer 
and one in the penultimate layer, at the critical frequency of the 
latter geometry, which is the lower of the two and lower than that
of the parent system. }
\label{removeE}
\end{figure}

An example application of the scale factor is for a system of ten 
chains of six molecules, where molecules are removed as the repulsion 
is increased.  The molecules are removed from the layer indicated by 
the degeneracy of the normal mode frequency that is going to zero at 
the critical repulsive frequency.  
It was found that the energetics correctly describe the order of the 
removal, which can be seen in figure \ref{removeE}.  When plotted relative 
to the ground state energy (and scaled down because the energies are quite 
large in our units), the point where the degeneracies switched to the
penultimate layer was also energetically lower than continuing to remove 
molecules from the outermost layer.  For the figure, the constants in the 
repulsive shift were those quoted above for the system of $(N,M)=(6,8)$, 
so one can see the mismatch for the removed molecules already appearing, 
as there should be no energy difference between the full system and the 
system where one molecule has been removed from the outermost layer at 
the critical frequency.  Thus if one is interested in detailed energetics, 
one should decide on a shift energy in the region where one is interested 
in doing calculations.

\section{Summary and Conclusion}\label{conclusionsect}
We have discussed how the harmonic approximation can be used to study the
$N$-body 2D problem of dipole-dipole interacting molecules.  First the
two-body interaction is approximated by a shifted harmonic oscillator
potential with the same negative and positive spatial regions. The
frequency is then determined as function of the molecular dipole
moment by the requirement that the two-body energy has to be
reproduced.  We test this procedure by comparing with the exact energy
of three and four molecules in separate layers. The approximation is remarkably
accurate over the range of available dipole moments.  The repulsion
between in-layer molecules is also approximated by a shifted harmonic
oscillator but now we leave the frequency as a parameter which can be 
related to properties of larger bound state structures when they become
available from exact calculations. 

The basic ingredient is a chain where one molecule is placed in each
layer where the intralayer repulsion then is absent. We calculate the
binding energy in units of the two-body energy of the number of
neighbouring pairs, and find a weak increase as function of molecule number.
The value is always only slightly larger than the analytical estimate
of $\zeta(3)$ which is approached for large dipole moment. Weaker
interactions result in about $10\%$ larger values reflecting more
correlation along the chain.  The nearest neighbours contribute by more
than $80\%$ of the energy. The spatial extension in the planes
measured in units of the layer distance from the total center of mass
position increases from central towards the outer layer. The weaker
the interaction the larger is the size. The normal modes are
oscillations away from equilibrium with a different number of nodes.  All
frequencies of the normal modes increase with interaction strength,
while the lowest of these always decrease with increasing molecule
number. 

The normal modes of the single chain configuration are equivalent to 
those obtained in the theory of acoustic phonons in crystals. By fitting
our numerical results for the lowest mode frequencies, one can obtain 
the speed of sound in the chain. We found that the speed of sound was 
more than a factor of three less than what one would naively get by
expanding the dipole potential around the origin to quadratic order. This 
reflects one of the differences of our procedure for obtaining an effective
harmonic approximation through the exact two-body bound state energy and 
the potential profile. The latter approach gives an effective more shallow 
potential which results in a smaller string tension and a reduced speed
of sound. This indicates a short-coming of the naive harmonic approach 
in describing the collective behaviour of a chain of dipoles.

We proceed to investigate two chains consisting of two molecules in
each layer.  The structure of the system is most clearly found through
its excitation modes, or specifically the normal modes. The repulsive
intralayer interaction is varied from zero to the critical value where
the system becomes unstable as seen by a vanishing normal mode
frequency.  The normal modes fall into two distinct groups.  The first
group is independent of the intralayer repulsion with a simple
one-to-one correspondence to the one-chain modes. These modes correspond
to in-phase motion of the two chains.
The energies of the
second group decrease with repulsion and the lowest of these modes vanishes at the
critical value. These modes correspond to in-layer oscillations
breaking the one-chain structure. This is analagous to optical 
modes in crystals with more than one atoms at each lattice site.
The breakup is seen through diverging
amplitudes on the molecules in the two outer layers.  

Increasing the number of chains so that there are more molecules in
each layer leads to remarkably similar behaviour of the normal mode
frequencies.  The one-to-one correspondence with one-chain modes
remains. Furthermore, the decreasing normal mode frequencies also
vanish at the same critical repulsion which then is independent of the
the number of molecules in the layers.  However, the degeneracy
increases in correspondence to the increase in the number of degrees of freedom 
within one
layer. This implies that the configuration at the breaking point has
diverging amplitudes of the molecules in the two outer layers.  The
breaking then leads to a system where the two outer layers have fewer
molecules than the more central layers. These configurations have less
repulsion and the next instability is related to removal of molecules
from the next two outer layers. This peeling off molecules would
continue to the third outermost layer or remove
more molecules in the outer layers. Eventually one chain only would
remain. We also calculate the density of the central layers
and find that it is independent of the number
of layers, depending on the number of molecules per layer to the power
$3/2$ with a factor that depends on the dipolar strength.

The structures are independent of the constant shifts in the 
repulsive potential terms which only have
influence on the relative energies. We choose these shifts such that
the energy of the system at the critical frequency is the same as the
most likely combination of fragments.
We adjust to relatively large
numbers of layers and molecules in each layer.  Then we compare
energies of a system and two systems of half the sizes either in
layers or molecules in the layers. We conclude that the more layers,
the more stable is the system, whereas the energy is rather
independent of the number of molecules in each layer.

In conclusion, the harmonic approximation is very useful in the
description of the two-dimensional systems of cold
polar molecules. However, this requires a careful adjustment of 
the forces between molecules in different planes and in 
the same plane. The latter can be particularly tricky to define
and we have discussed at length how to describe the in-plane
repulsion in a reasonable way using an inverted harmonic 
oscillator potential. Using a harmonic approximation for both
types of interaction is very desirable due to its exact solvability.
The structures are found as function of layers and
molecules in the layers, energies and sizes are calculated, and
configurations of excitation modes are found. The structures are
ordered in a hierarchy related to their stability. 

The results obtained 
here within the harmonic model indicate
that for a typical system where the strength of the inter-layer and 
intra-layer terms are both proportional to the square of the 
dipole moment, bound complexes with more than one molecules in a 
single layer are not bound and that the single chains with one 
molecule in each layer are the stable entities. However, the 
external confinement that any realistic experiment employs means that 
the dipoles cannot separate to large distances. We therefore expect that 
the multi-chain modes calculated in the harmonic model could show up 
as resonances. Here we provide a study of these modes and the 
breakup channels of multi-chain complexes.

The experimental conditions needed to study the systems considered here 
should be within reach of the next generation of cold polar molecular
experiments. The current front-runner uses $^{40}$K$^{87}$Rb molecules which have 
$U=0.1$ for the condition in Ref. \cite{miranda2011}, but which can probably
be extended up to about $U=1.2$ with increasing electric field. These values
of $U$ are likely too small for the harmonic approximation used here to 
apply since the two-body bound states are very extended \cite{armstrong2010}. 
However, most other combinations of alkali atoms into polar molecules can
potentially be used at dipole moments of 1 Debye or more \cite{julienne2011}, 
which would easily increase $U$ by one or two orders of magnitude. Larger 
values of $U$ implies that the single chain structures will have large binding
energies, and in turn be stable even in thermal samples. This means that one
does not necessarily need to be in the nano-Kelvin temperature regime
to study the chains discussed in the current work.

The current work indicates that the structure of a multi-layer system with
several particles in each layer is most likely that of single chains, 
i.e. the dipolar chains liquid originally proposed in Ref.~\cite{wang2006},
but that more complex states are possible if the intralayer repulsion could 
be somehow tuned independently of the interlayer attraction. This 
could perhaps be obtained by using a combination of external electric fields
and applied lasers \cite{giovanazzi2002,micheli2007,gorshkov2008}.

\end{document}